\documentclass[useAMS,usenatbib]{mnras}

\usepackage{threeparttable}
\usepackage{amsmath}
\usepackage{graphicx}
\usepackage{longtable}
\usepackage{amssymb}
\usepackage{tabularx, blindtext}
\usepackage{lscape}
\usepackage{lineno}
\usepackage{url}
\newcommand{\feh}{\ensuremath{{\rm [Fe/H]}}}
\newcommand{\teff}{\ensuremath{T_{\rm eff}}}
\newcommand{\teq}{\ensuremath{T_{\rm eq}}}
\newcommand{\logg}{\ensuremath{\log{g}}}
\newcommand{\zaspe}{\texttt{ZASPE}}
\newcommand{\ceres}{\texttt{CERES}}
\newcommand{\vsini}{\ensuremath{v \sin{i}}}
\newcommand{\kms}{\ensuremath{{\rm km\,s^{-1}}}}
\newcommand{\mjup}{\ensuremath{{\rm M_{J}}}}
\newcommand{\mearth}{\ensuremath{{\rm M}_{\oplus}}}
\newcommand{\mpl}{\ensuremath{{\rm M_P}}}
\newcommand{\rjup}{\ensuremath{{\rm R_J}}}
\newcommand{\rpl}{\ensuremath{{\rm R_P}}}
\newcommand{\rstar}{\ensuremath{{\rm R}_{\star}}}
\newcommand{\mstar}{\ensuremath{{\rm M}_{\star}}}
\newcommand{\lstar}{\ensuremath{{\rm L}_{\star}}}
\newcommand{\rsun}{\ensuremath{{\rm R}_{\odot}}}
\newcommand{\msun}{\ensuremath{{\rm M}_{\odot}}}
\newcommand{\lsun}{\ensuremath{{\rm L}_{\odot}}}

\newcommand{\mpkep}{\ensuremath{0.179 \pm 0.021 }}
\newcommand{\rpkep}{\ensuremath{0.840 \pm 0.011 }}
\newcommand{\mskep}{\ensuremath{1.105 \pm 0.019 }}
\newcommand{\rskep}{\ensuremath{1.669 \pm 0.022 }}
\newcommand{\per}{\ensuremath{11.63364 \pm 0.00010 }}
\newcommand{\ecc}{\ensuremath{0.420 \pm 0.034 }}

\newcommand{\plname}{EPIC~201498078b}
\newcommand{\stname}{EPIC~201498078}
\newcommand{\rhopl}{\ensuremath{{\rm \rho_P}}}
\newcommand{\rhopkep}{\ensuremath{0.37 \pm 0.05 }}
\newcommand{\gccm}{\ensuremath{\mathrm{g}\,\mathrm{cm}^{-3}}}
\newcommand{\taucirc}{\ensuremath{\tau_\mathrm{circ}}}

\renewcommand\vec[1]{\ensuremath{\mathbf{#1}}}

\title[\stname]{\plname: A low density Super Neptune
on an eccentric orbit}
\author[R. Brahm et al.]{R.\ Brahm,$^{1,2,3}$\thanks{E-mail: rbrahm@astro.puc.cl, Fondecyt Fellow}
                    N.\ Espinoza,$^{4,10}$                     
                    M.\ Rabus,$^{1,4}$  
                    A.\ Jord\'an,$^{2,3}$
                    M.~R.\ D\'iaz,$^{5}$ 
                    F.\ Rojas,$^{2}$ 
		            \newauthor
                    M.\ Vu\v{c}kovi\'{c},$^{7}$  
					A.\ Zapata,$^{1}$
                    C. Cort\'es$^{8,3}$
                    H.\ Drass,$^{1,3}$
					J.~S.\ Jenkins,$^{5}$
					R.\ Lachaume,$^{1,4}$
                    \newauthor
                    B.\ Pantoja,$^{5}$
                    P.\ Sarkis,$^{4}$
 					M.~G.\ Soto,$^{5}$                 
 					S.\ Vasquez,$^{9}$                        
                    Th.\ Henning,$^{4}$                    
                    M.~I.\ Jones,$^{6}$                    
\\
$^{1}$Center of Astro-Engineering UC, Pontificia Universidad Cat\'olica de Chile, Av. Vicu\~{n}a Mackenna 4860, 7820436 Macul, Santiago, Chile\\
$^{2}$Instituto de Astrof\'isica, Facultad de F\'isica, Pontificia Universidad Cat\'olica de Chile, Av. Vicu\~{n}a Mackenna 4860, 7820436 Macul,\\
Santiago, Chile\\
$^{3}$Millennium Institute of Astrophysics, Santiago, Chile\\
$^{4}$Max-Planck-Institut f\"{u}r Astronomie, K\"{o}nigstuhl 17, 69117 Heidelberg, Germany\\
$^{5}$Departamento de Astronom\'ia, Universidad de Chile, Camino El Observatorio 1515, Las Condes, Santiago, Chile\\
$^{6}$European Southern Observatory, Casilla 19001, Santiago, Chile\\
$^{7}$Instituto de F\'isica y Astronom\'ia, Universidad de Vapara\'iso, Casilla 5030, Valpara\'iso, Chile\\
$^{8}$Departamento de F\'isica, Facultad de Ciencias B\'asicas, Universidad  Metropolitana  de  la  Educaci\'on, Av. Jos\'e  Pedro Alessandri 774,\\
7760197, Nu\~noa, Santiago, Chile\\
$^{9}$Museo Interactivo Mirador, Dirección de Educación, Av. Punta Arenas 6711, La Granja, Santiago, Chile\\
$^{10}$Bernoulli Fellow\\
}

\begin{document}

\date{Draft Version 4.0}

\pagerange{\pageref{firstpage}--\pageref{lastpage}} \pubyear{2002}

\maketitle

\label{firstpage}

\begin{abstract}
We report the discovery of \plname, which was first identified
as a planetary candidate from \textit{Kepler} K2 photometry of Campaign 14,
and whose planetary nature and orbital parameters were then confirmed
with precision radial velocities.
\plname\ is half as massive as Saturn ($\mpl = \mpkep\,\mjup$),
and has a radius of $\rpl= \rpkep\,\rjup$, which translates into a bulk density of
$\rhopl =\rhopkep\,\gccm$.
\plname\ transits its slightly evolved G-type host star ($\mstar = \mskep\,\msun$, $\rstar = \rskep\,\rsun$) every \per\ days and presents a significantly eccentric orbit ($e=\ecc$).
We estimate a relatively short circularization timescale of 1.8\,Gyr for the planet, but
given the advanced age of the system we expect the planet to be engulfed by its evolving
host star in $\sim 1$\,Gyr before the orbit circularizes. 
The low density of the planet coupled to the brightness of the host star
($J=9.4$) makes this system one of the best candidates known to date in the
super-Neptune regime for atmospheric characterization via transmission spectroscopy, and to further study the transition region between ice and gas giant planets.

\end{abstract}


\section{Introduction}
\label{sec:intro}

Transiting planets in the transition region between ice- and gas-giants \citep[super Neptunes, e.g.][]{hats8b} are fundamental objects for constraining planet formation theories,
particularly regarding the runaway accretion of the gaseous
envelope \citep{ida:2004,mordasini:2009}.
The physical parameters (mass and radius) of these planets can be
compared to theoretical models in order to infer their internal composition
(heavy element content), which can then be linked to the observed stellar
properties and predicted proto-planetary disc conditions \citep{thorngren:2016}.
In addition, the detection of molecules in the atmospheres of these planets
via transmission spectroscopy can be connected to different orbital distances at
which the formation of the planet or accretion of the envelope took place
\citep{mordasini:2016,Espinoza:2017}.
Additionally, detailed characterization of the orbital parameters of these
systems can deliver clues about their formation and/or migration histories
\citep{dawson:2013,petrovich:2016}.

Nonetheless, super Neptunes are among the least studied type of transiting planets to date, due to the low number of detected systems around bright stars. The photometric precision of
ground-based  surveys is barely enough to discover planets slightly
smaller than Jupiter \citep{hatp11,demangeon,hats43-46}, and hence most of what we know about these planets has come from statistical studies using ground-based radial velocity observations \citep[e.g.][]{jenkins17,udry:2017}.
On the other hand, most of the Kepler systems found in the super Neptune region
are too faint for determining their masses via precision radial velocities.
This picture has started to change thanks to the $Kepler$ K2 mission \citep{howell:2014}, which
has been able to monitor a significant number of bright stars with very
high photometric precision, resulting in the discoveries of several of these types of planets \citep[e.g.][]{vaneylen,k255,k298}. We expect this number to grow significantly once the first light curves from
the TESS mission become available \citep{tess}.

In this study we present the discovery of a warm super Neptune planet
that transits its host star, and presents one of the largest eccentricities
known for short period planets in this mass range.
This discovery was performed in the context of the K2CL collaboration,
which uses spectroscopic facilities located in Chile to confirm
and characterize transiting planets from K2 \citep{brahm:2016:k2,espinoza:2017:k2,jones:2017,giles:2018,soto:2018,k2-232}.
The structure of the paper is as follows. In \S~\ref{sec:obs} we present the
photometric and spectroscopic observations that allowed the discovery of \plname,
in \S~\ref{sec:ana} we derive the planetary and stellar parameters,
and we discuss our findings in \S~\ref{sec:disc}. 

\section[]{Observations}
\label{sec:obs}

\subsection{Kepler K2}
\label{sec:k2}
\begin{figure*}
 \includegraphics[width=2\columnwidth]{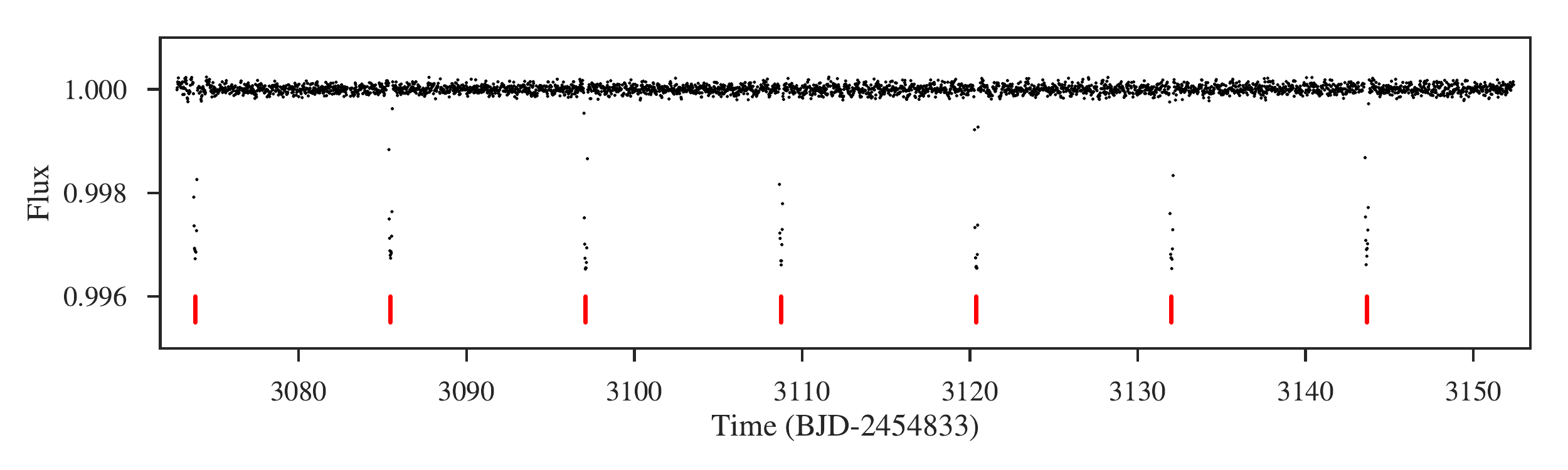}
 \caption{De-trended K2 photometry of \stname. The positions of the transits of our planet candidate are highlighted with red marks.}
 \label{fig:photometry}
\end{figure*}

Observations of field 14 of K2 mission ($\mathrm{RA}=10^\mathrm{h}42^\mathrm{m}44^\mathrm{s}$, $\mathrm{DEC} = +06^\mathrm{o}51^\prime06^{\prime\prime}$) were performed between May and August of 2017, and were released to the community on November of the same year.
The photometric data for all targets of this campaign was reduced from pixel-level products to photometric light curves using the \texttt{EVEREST} algorithm \citep{EVEREST1,EVEREST2}, where long-term trends in the data were corrected with a Gaussian-process regression.
As in previous K2 campaigns, the transiting planet detection was performed by using the Box-fitting Least Squares algorithm \citep{BLS} on the processed light curves.
With this procedure we identified 41 planetary candidates, among which
\stname\ with an estimated period of 11.6 day, was catalogued as high
priority due to the brightness of the star and its clean box-shaped
transits with depths of $\approx 3000$\,ppm (see Figure~\ref{fig:photometry}).

\subsection{Spectroscopic Observation}

\label{sec:spec}
Our spectroscopic follow-up campaign consists primarily in the use of
four stabilized echelle spectrographs installed at the ESO La Silla
Observatory in Chile. We use the FIDEOS spectrograph installed on the ESO\ 1\,m telescope \citep{vanzi:2018} and the Coralie spectrograph \cite{coralie:1998} on the 1.2\,m-Euler/Swiss telescope for performing an initial characterization of the host star in order to
reject false positives. Particularly for the case of \stname, we obtained
three Coralie spectra on three different nights between January and March of 2018. Observations
were performed with the simultaneous calibration mode \citep{baranne:96} in which the comparison
fibre is illuminated with a Fabry-P\'erot etalon \citep{coralieFPI} in order to trace the instrumental
drift produced by environmental changes during the observation. The adopted exposure times ranged between 1500 and 1800\,s, and achieved a signal-to-noise
ratio of 40--50 per resolution element. These three spectra were reduced
and processed with the automated \ceres\ pipeline \citep{jordan:2014,brahm:2017:ceres}, which performs the optimal
extraction and wavelength calibration, and delivers precision radial velocities, bisector span measurements, and an estimation of the atmospheric parameters.
No additional stellar components were identified in the spectra, and no large amplitude velocity variations were observed, which could have been originated by
an eclipsing stellar mass companion. 

\begin{figure*}
 \includegraphics[width=\textwidth]{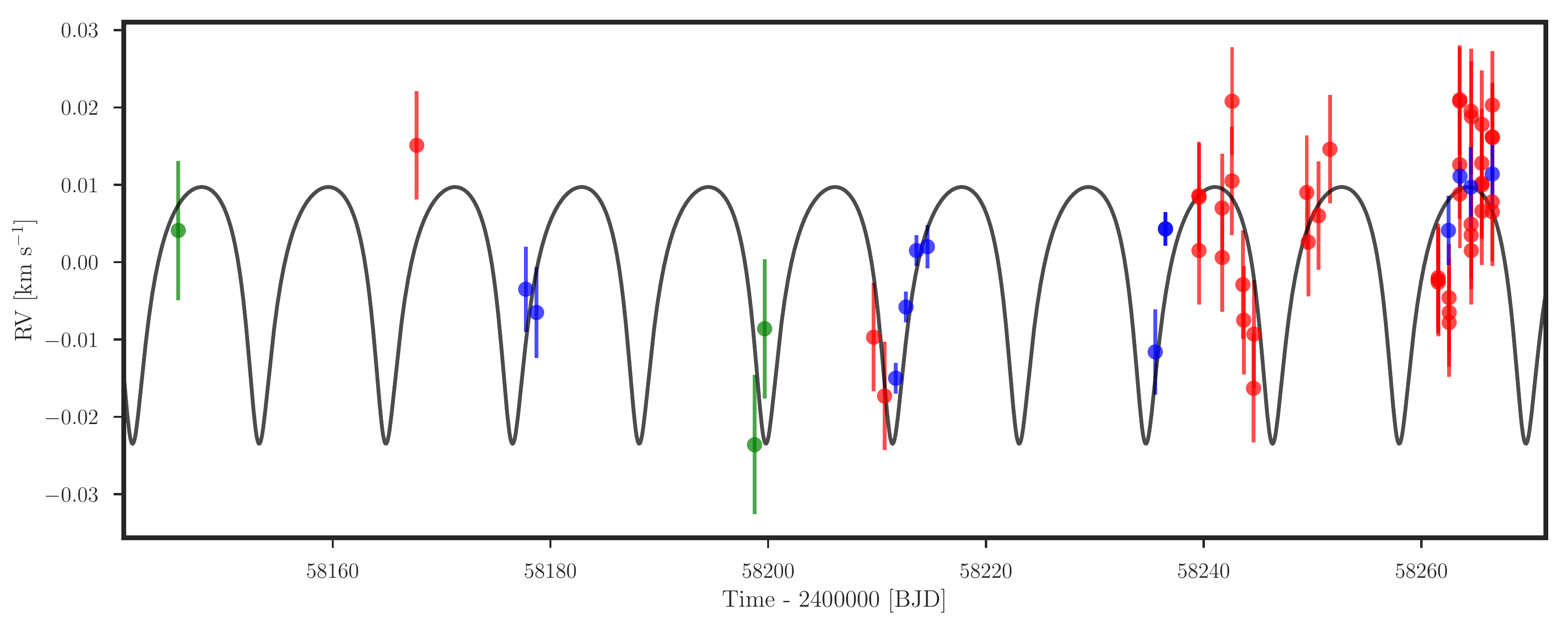}
 \caption{ Radial velocity (RV) curve obtained with Coralie (green), FEROS (red) and HARPS (blue). The black line corresponds to the Keplerian model with the posterior parameters found in Section 3. }
 \label{rvs-t}
\end{figure*}

After this initial characterization, more powerful facilities are required
to determine the mass and orbital parameters of the hypothetical planetary companion.
We acquired 43 spectra with FEROS \citep{kaufer:99} installed on the MPG2.2m telescope and 12
spectra with HARPS \citep{mayor:2003} installed on the ESO3.6m telescope. These observations
were performed between March and May of 2018. In the case of the FEROS observations, the instrumental drift during the science observations was monitored with the secondary fibre which was illuminated by a ThAr+Ne lamp. No instrumental drift was monitored during the HARPS observations because the stability of
this instrument is significantly higher than the amplitude in radial velocity that we intend to measure. The
comparison fibre was pointed to the background sky. Table \ref{tab:specobs} summarizes the general properties of these observations.
The data from both instruments was processed through the \ceres\ package in order to obtain precision radial velocities and bisector span measurements from the raw science images. These measurements are presented in Table~\ref{rvsbss}, and the
radial velocity curve is shown in Figure~\ref{rvs-t}.

\begin{figure}
 \includegraphics[width=\columnwidth]{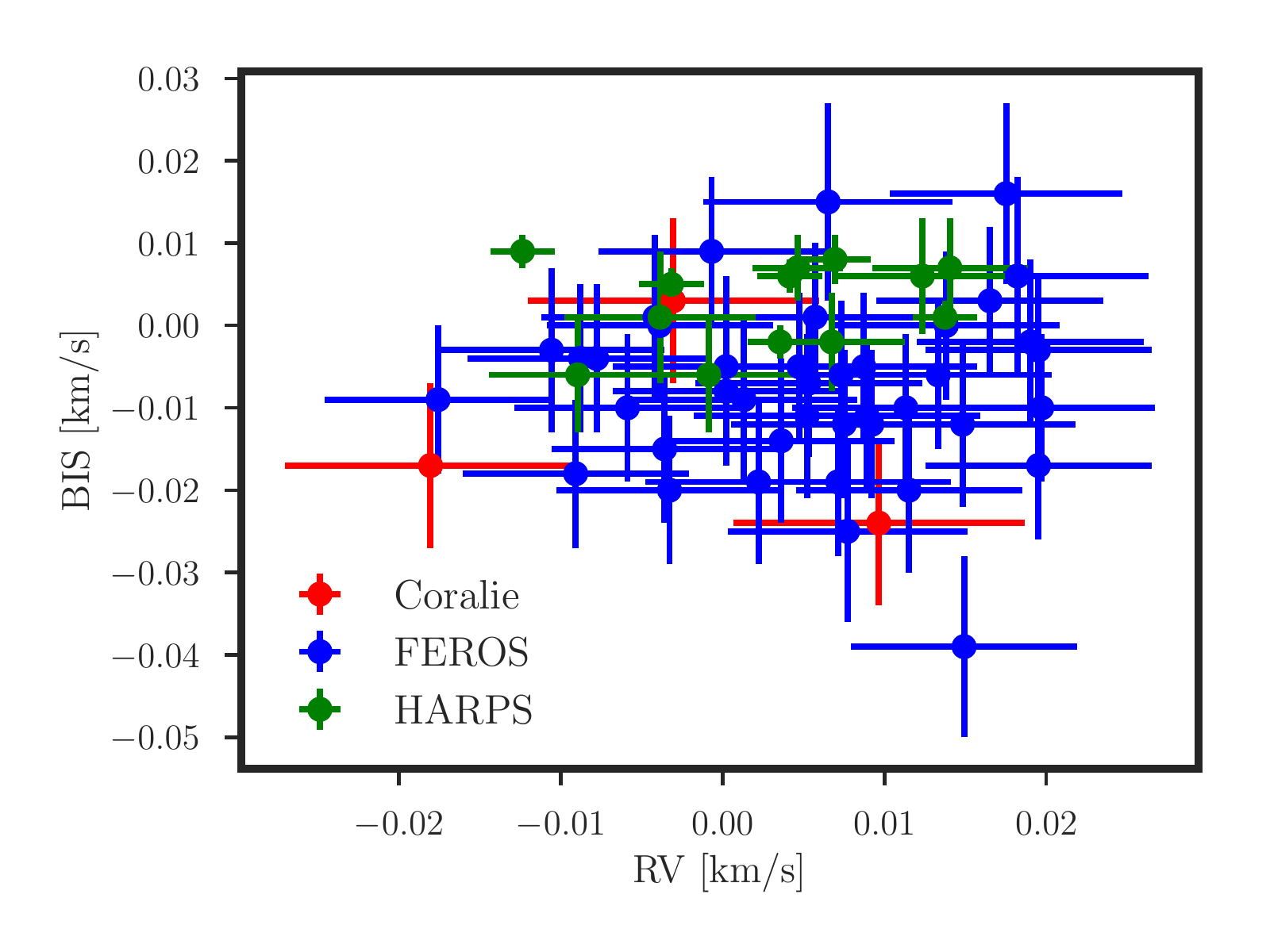}\\
 \caption{ Radial velocity (RV) versus bisector span (BIS) scatter plot using data from our spectroscopic observations of \stname. No significant correlation was found. }
 \label{rvs-bs}
\end{figure}

The velocities present a periodic variation that matches that of the transit signal.  In fact, a blind and wide-parameter search performed by the \texttt{EMPEROR} code \citep{jenkins18} on the RVs uncovered a statistically significant signal matching the properties indicated by the transit data, providing independent confirmation of the reality of the planet.  The relatively low amplitude of the variation implies a sub Jovian mass for the orbital companion, while the shape of the variation shows that the orbit is significantly non-circular. Additionally, a scatter plot of the radial velocities and bisector span values is displayed in Figure~\ref{rvs-bs}. No significant correlation is observed, which further reduces the probability that the system corresponds to a blended eclipsing binary or that the radial velocity signal is produced by stellar activity. In concrete terms, the median \textit{Pearson} coefficient value between RVs and BIS is $-0.15$ with a 95\% confidence interval between $-0.44$ and $0.24$, representing no statistically significant correlation.
Finally, the stellar atmospheric parameters reported by \texttt{CERES}
in the case of the three instruments were consistent with those of a subgiant star ($\teff \approx 5400$\,K, $\logg\approx3.8$, $\feh \approx 0.2$).

\begin{table*}
 \centering
 \begin{center}
  \caption{Summary of spectroscopic observations for \stname.}
  \begin{tabular}{cccccc}
  \hline
   Instrument     &   UT Date(s)  & N Spec. &   Resolution &  S$/$N range  &  RV Precision [m s$^{-1}$] \\
 \hline
 Coralie / 1.2m Euler/Swiss   & 2018 Jan -- Mar  &  3 &  60000 &  27 --  44  & 10 \\
 FEROS / 2.2m MPG             & 2018 Jan -- May &  43 &  50000 & 106 -- 167  &  7 \\
 HARPS / 3.6m ESO             & 2018 Jan -- May &  12 & 115000 &  36 --  50  &  2 \\
\hline
\end{tabular}
 \label{tab:specobs}
 \end{center}
\end{table*}

\subsection{GAIA}
\stname\ was also observed by GAIA. According to GAIA DR2 \citep{dr2}, no companions were identified in the neighbourhood of the star, which could be affecting the transit depth measured with the $Kepler$ telescope \citep{evans:2018}. Additionally, the reported radial velocity ($3.28 \pm 0.52\,\kms$) and \teff\ ($5288 \pm 120$\,K) values are consistent with those obtained through our spectroscopic follow-up observations. A parallax of $\pi = 4.660 \pm 0.043$\,mas is reported, which we use to determine the stellar properties as described in Section \ref{sec:ana}. 

\section{Analysis}
\label{sec:ana}
\subsection{Stellar parameters}
\label{sec:stellar-parameters}
For computing the physical parameters of \stname\ we follow an iterative procedure that consists in the following steps:

\begin{itemize}
\item Determination of the stellar atmospheric parameters (\teff, \logg, [Fe/H], \vsini) from high resolution spectra.
\item Determination of the stellar radius (\rstar) and extinction factor ($A_V$) from the GAIA parallax and available photometry.
\item Determination of the stellar mass (\mstar) and age by comparing the obtained stellar radius and \teff\ with theoretical evolutionary tracks.
\item Computation of a new \logg\ value from \rstar\ and \mstar, which is used as a fixed parameter in the following iteration.
\end{itemize}

Particularly for the case of \stname\ only two iterations were required. We obtained the atmospheric parameters with the \zaspe\ code \citep{brahm:2015,brahm:2016:zaspe} from the co-added HARPS spectra.
\zaspe\ determines \teff, \logg, \feh, and \vsini\ by
comparing the observed spectrum to synthetic ones in the spectral regions most sensitive to changes in those parameters. Additionally, reliable uncertainties are obtained from the data by performing Monte Carlo simulations that take into account the systematic mismatches between data and models.
Using this procedure we obtained the following parameters for the first iteration:
 $\teff = 5532 \pm 70$\,K, $\logg = 4.04 \pm 0.127$\,dex, $\feh = 0.26 \pm 0.04$\,dex, and $\vsini = 3.44 \pm 0.43\,\kms$, which were close to
 those obtained in the final iteration: 
  $\teff = 5513 \pm 51$\,K, $\logg = 4.032 \pm 0.015$\,dex, $\feh = 0.26 \pm 0.03$\,dex, and $\vsini = 3.7 \pm 0.17\,\kms$. Additionally, we performed an independent estimation of the stellar atmospheric parameters using the \texttt{SPECIES} \citep{species} finding consistent results at the 1$\sigma$ level to those computed with \zaspe. 
  
\begin{figure}
 \includegraphics[width=\columnwidth]{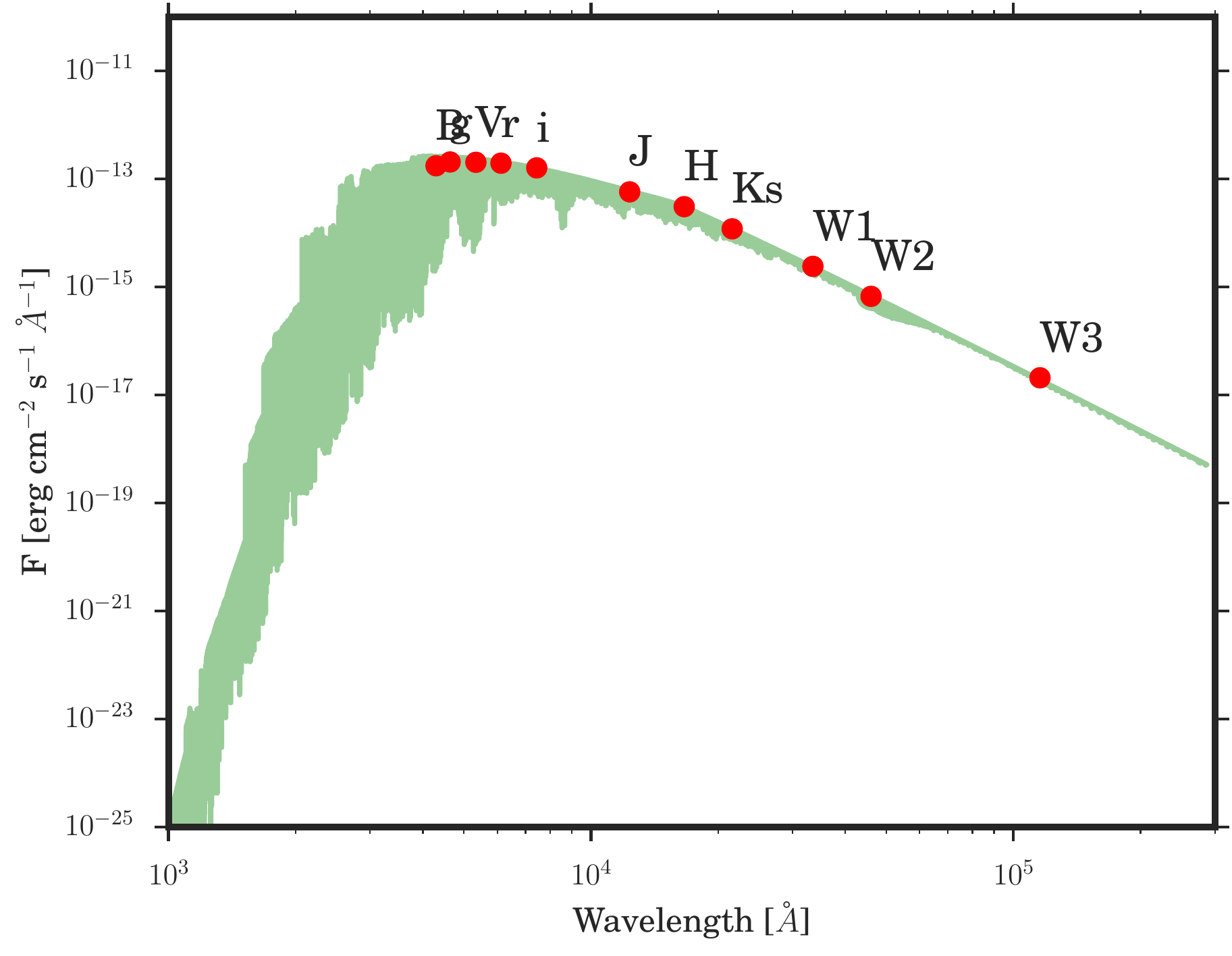}
 \caption{Spectral energy distribution of the \texttt{BT-Settl-CIFIST} model for the atmospheric parameters of \stname. The red circles correspond to the observed flux densities for the different passband filters are identified as red circles.}
 \label{sed}
\end{figure}

For the second step we used the \texttt{BT-Settl-CIFIST} spectral models from \citet{baraffe:2015}, which were interpolated to generate a synthetic spectral energy distribution (SED) consistent with the atmospheric parameters of \stname.
We then integrated the SED in different spectral regions to generate synthetic magnitudes that were weighted by the corresponding
transmission functions of the passband filters. 
The synthetic SED along with the observed flux density  in the different filters are plotted in Figure \ref{sed}.

Following the same procedure described in \citet{k2-232}, these synthetic magnitudes were used to infer the stellar radius (\rstar) and the extinction factor ($A_V$) by comparing them to the observed
magnitudes after applying a correction of the dilution of the stellar flux due to the distance computed from the GAIA parallax. 
We used the \texttt{emcee} \texttt{Python} package \citep{emcee:2013} to sample the posterior distribution of  R$_{\star}$ and $A_V$.
We also repeated this process for different values of \teff\ sampled from a Gaussian distribution, finding that the uncertainty in the final stellar radius is dominated by the uncertainty in \teff.
The stellar radius obtained in the last iteration was of \rstar=\rskep\ \rsun, which includes the error budget provided by the uncertainty in \teff.

\begin{figure}
 \includegraphics[width=\columnwidth]{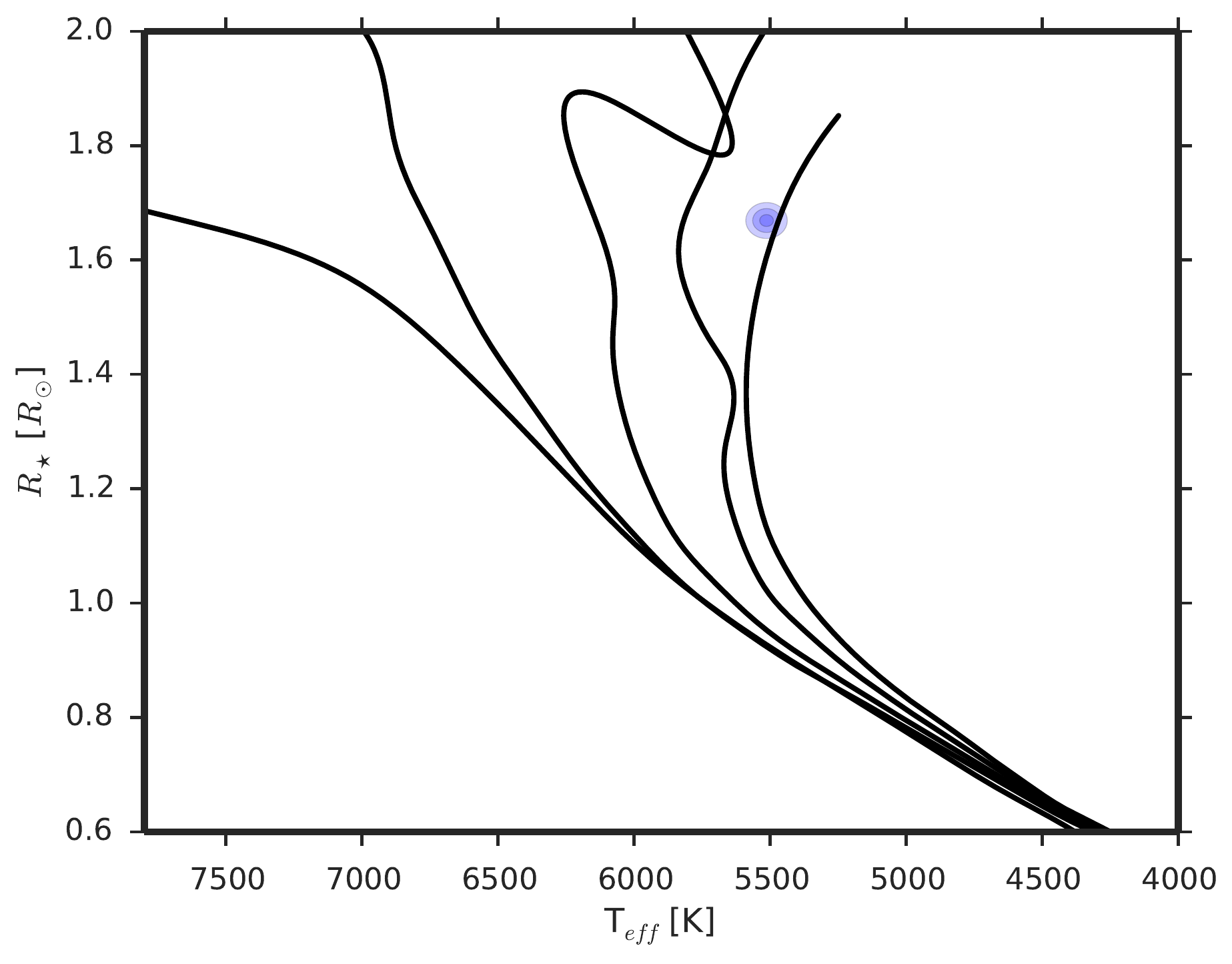}
 \caption{Yonsei-Yale isochrones for the metallicity of \stname\ in the \teff--\rstar\ plane. From left to right the isochrones correspond to 0.1, 1, 3, 6, 9 Gyr. The position of \stname\ in this plane is shown in blue, where the three different intensities represent the 1$\sigma$, 2$\sigma$, and 3$\sigma$ contours.}
 \label{iso}
\end{figure}

For the third step we used the Yonsei-Yale isochrones \citep{yi:2001}, which were interpolated
to the metallicity found with \zaspe.
We used the \texttt{emcee} package to explore the
posterior distribution of the stellar masses and
ages that generate stellar radii and effective temperatures consistent with the values found in the previous steps.
Figure~\ref{iso} shows
different isochrones in the  \teff--\rstar{} plane, with the value for \stname\ indicated in blue. 
The final parameters obtained for \stname\ are listed in Table~\ref{tab:stellar}.
We found that \stname\ is a G-type star with a mass of $\mstar=\mskep\,\msun$, that has just recently
started to depart from the main sequence at an age of $8.5 \pm 0.5$\,Gyr.

\begin{table}
 \centering
  \caption{Stellar properties and parameters for \stname.}
  \label{tab:stellar}
  \begin{tabular}{@{}lcc@{}}
  \hline
   
  Parameter      &   Value &  Method / Source \\
 \hline
 \\
Names     & \stname\ & -- \\
RA            & $10^\mathrm{h}52^\mathrm{m}07.78^\mathrm{s}$ & -- \\
DEC           & $00^\text{o}29^\prime36.07^{\prime\prime}$ &   \\
Parallax  [mas]    & 4.660 $\pm$ 0.042 & GAIA\\
\hline
\\
$K_p$ (mag) & 10.451 & EPIC\\
$B$ (mag) &11.561 $\pm$ 0.086 & APASS\\
$g$ (mag) & 10.979 $\pm$  0.010 & APASS\\
$V$ (mag) & 10.612 $\pm$  0.059 & APASS\\
$r$ (mag) & 10.402 $\pm$  0.020 & APASS\\
$i$ (mag) & 10.226 $\pm$  0.020  & APASS \\
$J$ (mag) & 9.337 $\pm$ 0.030 & 2MASS\\
$H$ (mag) & 8.920 $\pm$ 0.042 & 2MASS\\
$K_\mathrm{s}$ (mag) & 8.890 $\pm$ 0.022& 2MASS\\
W1 (mag) & 8.828 $\pm$0.023& WISE\\
W2 (mag) & 8.897 $\pm$ 0.020& WISE\\
W3 (mag) & 8.819 $\pm$ 0.031& WISE\\
\hline
\\
\teff\  [K] &  5513 $\pm$ 51 & \zaspe \\
\logg\ [dex]     &  4.032 $\pm$ 0.015 & \zaspe \\
\feh\ [dex]    &  0.26 $\pm$ 0.03 & \zaspe \\
\vsini\ [\kms] & 3.70 $\pm$ 0.17 & \zaspe \\
\hline
\\
\mstar{} [\msun] &  $1.104_{-0.018}^{+0.020}$ & \zaspe\ + GAIA + YY \\
\rstar{} [\rsun] & $1.669_{-0.021}^{+0.022}$ & \zaspe\ + GAIA  \\
\lstar{}  [\lsun] & $2.32_{-0.11}^{+0.12}$ & \zaspe\ + GAIA  + YY\\
Age    [Gyr]        &    $8.51_{-0.48}^{+0.53}$ & \zaspe\ + GAIA + YY \\
$A_V$  &  $0.030_{-0.017}^{+0.017}$ & \zaspe\ + GAIA  \\
\hline

\end{tabular}
\end{table}

\subsection{Global modeling}
In order to determine the orbital and transit parameters of the \plname\ 
system we performed a joint analysis of the de-trended photometry of the
\textit{Kepler} telescope and follow-up radial velocities.
As in previous planet discoveries of the K2CL collaboration, we
used the \texttt{exonailer} code which is described in
detail in \citet{espinoza:2016:exo}. Briefly, we model the transit light
curves using the \texttt{batman} package \citep{kreidberg:2015} by taking
into account the smearing effect of the transit shape produced by the
long-cadence of K2 \citep{ kipping:2010}.
To avoid systematic biases in the determination of the transit parameters
we considered the limb-darkening coefficients as additional free parameters in the transit modelling \citep{EJ:2015}, where the limb-darkening law to use is decided following \citet{espinoza:2016:lds}; in our case, we select the quadratic limb-darkening law. The limb-darkening coefficients were fit using the 
uninformative sampling technique of \citet{Kipping:LDs}. We also include a 
photometric jitter parameter, which allow us to have an estimation of the level of stellar noise in the light curves. As for the radial velocities, they are modelled with
the \texttt{rad-vel} package \citep{fulton:2018}, where we consider a
different systemic velocity and jitter factor for the data of each spectrograph. 
In addition, we also used the stellar density derived in our stellar modelling as an extra ``data point" in our global fit, with the idea being that, given $\vec{y}=\{\vec{y}_{tr}, \vec{y}_{rv}, \vec{y}_{\rho_*}\}$ is the data vector containing the transit data $\vec{y}_{tr}$, the radial-velocity data $\vec{y}_{rv}$ and the 
stellar density ``data" (obtained form our stellar analysis using GAIA+Isochrones) $\vec{y}_{\rho_*}$, because the three pieces of data are independent, the likelihood can be decomposed as $p(\vec{y}|\theta ) = p(\vec{y}_{tr}|\theta )p(\vec{y}_{rv}|\theta )p(\vec{y}_{\rho_*}|\theta )$. The latter term is assumed to be Gaussian, and is 
given by
\begin{eqnarray*}
p(\vec{y}_{\rho_*}|\theta ) = \frac{1}{\sqrt{2\pi \sigma^2_{\rho_*}}} \exp -\frac{(\rho_* - \rho_*^m)^2}
{2\sigma_{\rho_*}^2} ,
\end{eqnarray*}
where 
\begin{eqnarray*}
   \rho^m_* = \frac{3\pi}{GP^2}\left(\frac{a}{R_*}\right)^3
\end{eqnarray*}
by Kepler's law, and 
$\rho_*$ and $\sigma_{\rho_*}$ are the mean stellar density and its standard-deviation,
respectively, derived from our stellar analysis. In essence, because the period $P$ is 
tightly constrained by the observed periodic transits, this extra term puts a strong 
constraint on $a/R_*$, which in turn helps to extract information about the eccentricity $e$ and argument of periastron $\omega$ from the duration of the transit. This 
methodology has been updated in \texttt{exonailer}. Both an eccentric and a 
circular model were considered, but the eccentric model was the one favoured by the 
data (the Bayasian  information criterium in favour of the latter model with $\Delta \textnormal{BIC} = 5.4$).

\begin{figure}
 \includegraphics[width=\columnwidth]{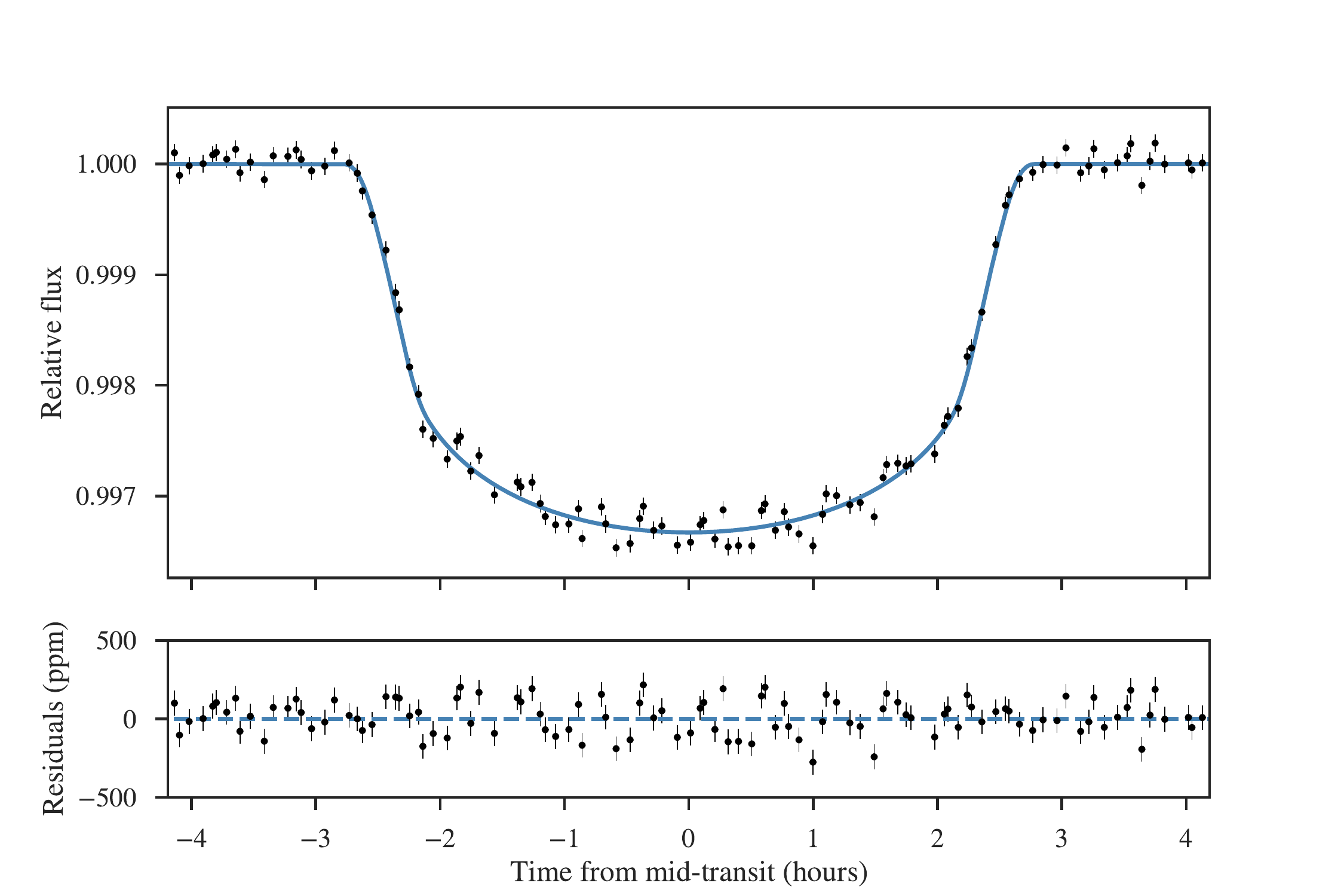}
 \caption{ The top panel shows the phase folded Kepler $K2$ photometry (black points) for the \stname\ system around the transit,
 and the model generated with the derived parameters of \texttt{EXONAILER} (blue line). The bottom panel shows the corresponding residuals.}
 \label{exonailerlc}
\end{figure}

\begin{figure}
 \includegraphics[width=\columnwidth]{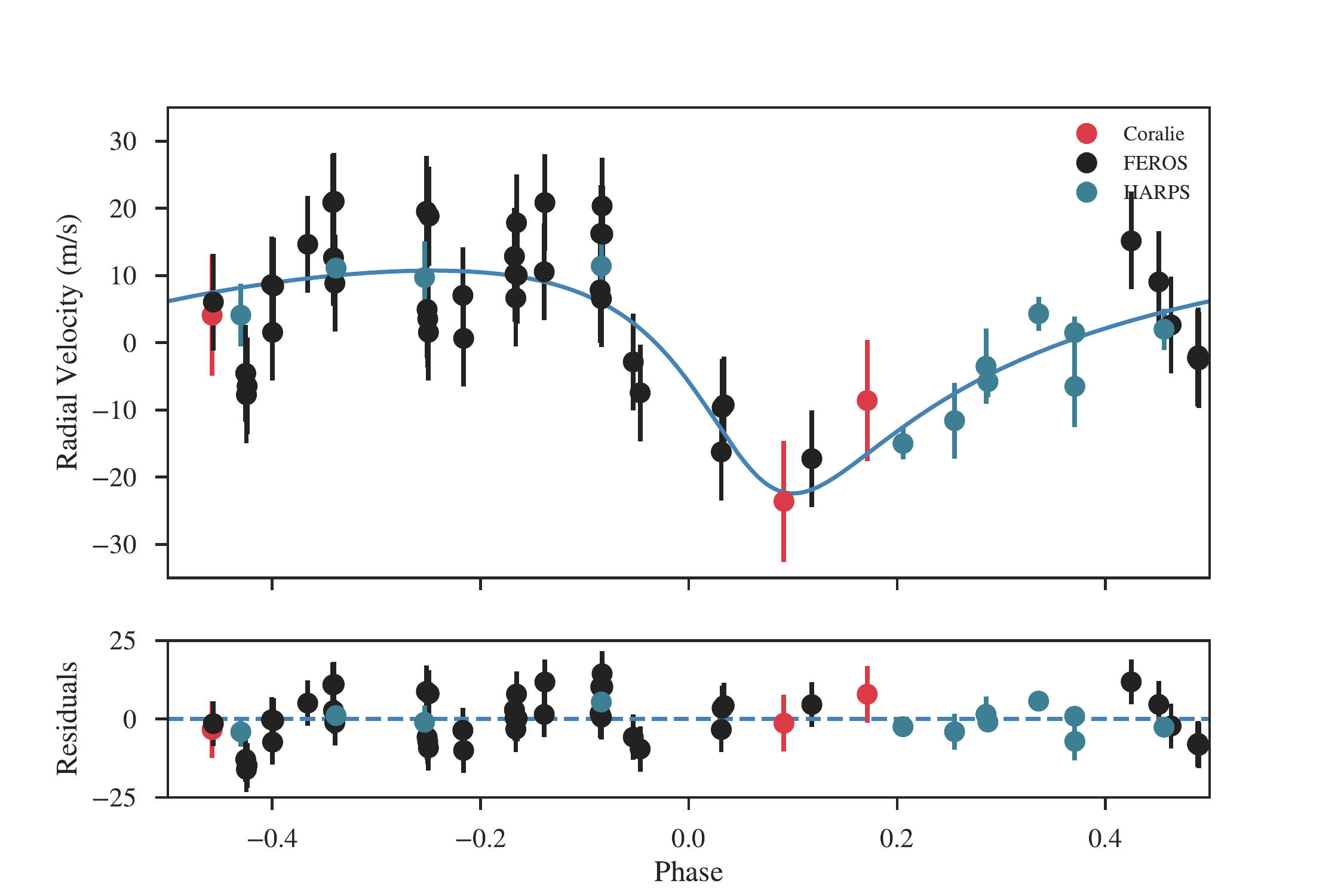}
 \caption{ The top panel presents the radial velocities (coloured circles) obtained with the three spectrographs as a function of the orbital phase.
 The RV model with the derived orbital parameters for \plname corresponds to the blue solid line. The bottom panel shows the residuals obtained
 for these radial velocity measurements.
 }  
 \label{exonailerrv}
\end{figure}

The adopted priors and obtained posteriors of our modelling are presented in Table~\ref{tab:planet}.
Figures \ref{exonailerlc} and \ref{exonailerrv} show the adopted solution
for the transit and orbital variation of \plname\ as a function of the orbital phase, generated from the posterior distributions.
The corresponding data is also presented in this plot.
We combined the inferred stellar physical properties of \stname\ with the 
obtained transit and orbital parameters to obtain the physical parameters
of the planet.
We found that the mass of \plname\ lies in the super Neptune regime ($\mpl=\mpkep\,\mjup$) while its radius is slightly larger that the one of Saturn ($\rpl=\rpkep\,\rjup$).
These values imply a relatively low bulk density of $\rhopl=\rhopkep\,\gccm$. Additionally, due to its moderately long orbital period the nominal equilibrium temperature of \plname\ is of $\teq = 1065 \pm 12$\,K. However, due to the high eccentricity of the system ($e = 0.42 \pm 0.03$), the atmospheric
temperature of \plname\ could be significantly different depending on the
atmospheric circulation properties of the planet.

\begin{table*}
 \centering
 \begin{center}
   \caption{Transit, orbital, and physical parameters of \plname. On the priors, $N(\mu,\sigma)$ stands for a normal distribution with mean $\mu$ and standard deviation $\sigma$, $U(a,b)$ stands for a uniform distribution between $a$ and $b$, and $J(a,b)$ stands for a Jeffrey's prior defined between $a$ and $b$.}
   \label{tab:planet}
 \begin{threeparttable}

  \begin{tabular}{@{}lcc@{}}
  \hline
   
  Parameter      &   Prior &  Value \\
 \hline
 Light-curve parameters \\
 
$P$ (days) & $N(11.63362,0.01)$ & $11.63365 \pm 0.00010$\\
$T_0$ (days) & $N(2457906.84,0.01)$ & $2457906.83957 \pm 0.00049$\\
$\rpl/\rstar$ & $U(0.001,0.5)$ & $0.05178^{+0.0006}_{-0.0005}$\\
$a/\rstar$ & $U(1,30)$ & $13.38^{+0.16}_{-0.18}$\\
$i$ & $U(0,90)$  & $89.20^{+0.57}_{-0.68}$ \\
q$_1$ & $U(0,1)$ & $0.48^{+0.11}_{-0.09}$\\
q$_2$ & $U(0,1)$ & $0.36^{+0.08}_{-0.07}$\\
$\sigma_w$ (ppm) & $J(10,5000)$ & $79.03^{+0.99}_{-0.95}$\\

\hline
RV parameters\\

K  (m s$^{-1}$)  &  $N(0,100)$ & $16.6^{+1.9}_{-1.9}$\\
$e$              &  $U(0,1)$    & 0.42 $\pm$ 0.03 \\
$\omega$ (deg)   &  $U(0,360)$  & 147.0 $^{+5.9}_{-6.3} $\\
$\gamma_\mathrm{coralie}$ (m s$^{-1}$) & $N(3300,100)$ &   $3328.1^{+6.1}_{-5.9} $ \\
$\gamma_\mathrm{feros}$ (m s$^{-1}$)   & $N(3300,100)$ &   $3319.9^{+1.3}_{-1.4} $\\
$\gamma_\mathrm{harps}$ (m s$^{-1}$)   & $N(3300,100)$ &   $3341.7^{+1.4}_{-1.5} $ \\
$\sigma_\mathrm{coralie}$ (m s$^{-1}$) & $J(10^{-2},10)$ &  $0.7^{0.8}_{-0.8} $ \\
$\sigma_\mathrm{feros}$ (m s$^{-1}$)   & $J(10^{-2},10)$ &  $1.7^{+2.8}_{-0.9} $ \\
$\sigma_\mathrm{harps}$ (m s$^{-1}$)   & $J(10^{-2},10)$ &  $2.2^{+2.1}_{-1.1} $ \\

\hline
Derived parameters\\

\mpl\ (\mjup)     & -- & $0.179 \pm 0.020$ \\
\rpl\ (\rjup)      & -- & $0.840 \pm 0.011$ \\
\teq\ (K)         & -- & $1066_{-13}^{+11}$   \\
$a$  (AU)           &  -- & $0.10376_{-0.00051}^{+0.00043}$   \\
\rhopl\ (\gccm) &  -- & $0.37 \pm 0.05$   \\
\hline

\end{tabular}
  \end{threeparttable}
 \end{center}
\end{table*}

\subsection{Rotational modulation and search of additional transits}
The data with masked transits were used in order to search for rotational modulations from the star and/or the planet, as well 
as further possible transits and/or secondary eclipses in the data. 
The detrended data did not show any indication of secondary 
eclipses and/or phase curve modulations, which was expected 
as the secondary eclipse depth is, at maximum, of order $(\rpl/a)^2 = 15.40^{+0.58}_{-0.53}$\,ppm if only a reflected light component is 
considered, which is five times below the attained noise level in 
the K2 photometry. As for further transits, a BLS on the data 
reveals two prominent peaks at $P=0.81$ and $P=7.39$ days, yet 
no clear transit signature at those periods is observed. No 
rotational modulation signatures were detected in the photometry.
\section{Discussion}
\label{sec:disc}

\begin{figure*}
 \includegraphics[width=\textwidth]{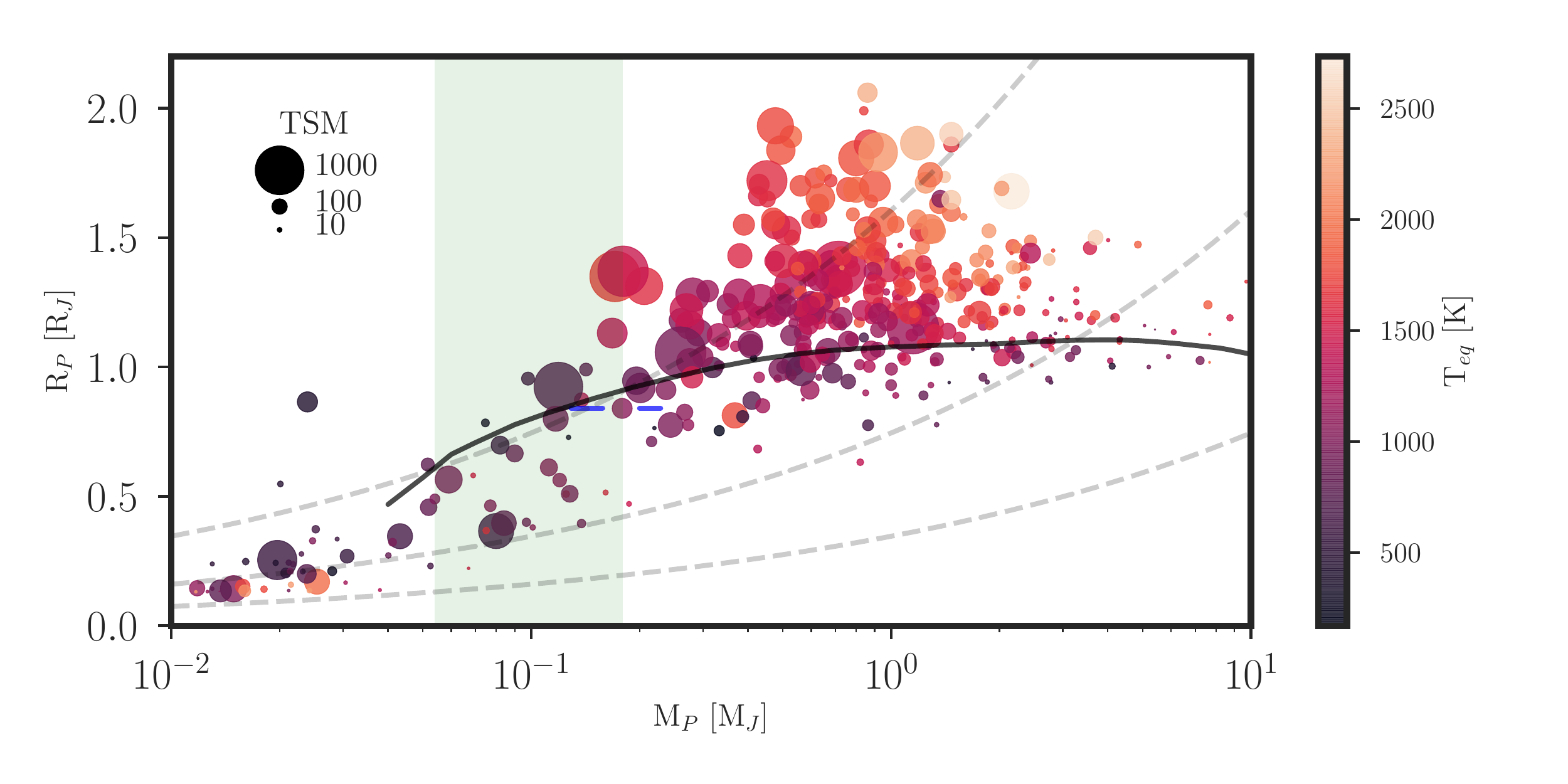}
 \caption{Mass--Radius diagram for the population of transiting planets
 with masses and radii measured with a precision better than 25\%. The points
 are colour coded according to their equilibrium temperature. The size of the
 points scales with the transmission spectroscopy metric as defined by \citep{tsm}.
 The continuous black line corresponds to the theoretical radius
 of the \citet{fortney:2007} models for a core mass of 10\mearth and
 a similar insolation than \plname. The dashed lines correspond to different iso-density curves of 0.3, 3, and 30,\gccm, from top to bottom.
 The super Neptune region, which corresponds  to the ice-gas transition of giant planets is highlighted in green. \plname\ (blue error bars) falls in the region of moderately irradiated low density super Neptunes and its structure would be consistent to that of a
10\mearth ice-rocky core of surrounded by a solar metallicity envelope. }
 \label{m-r}
\end{figure*}

\begin{figure*}
 \includegraphics[width=\textwidth]{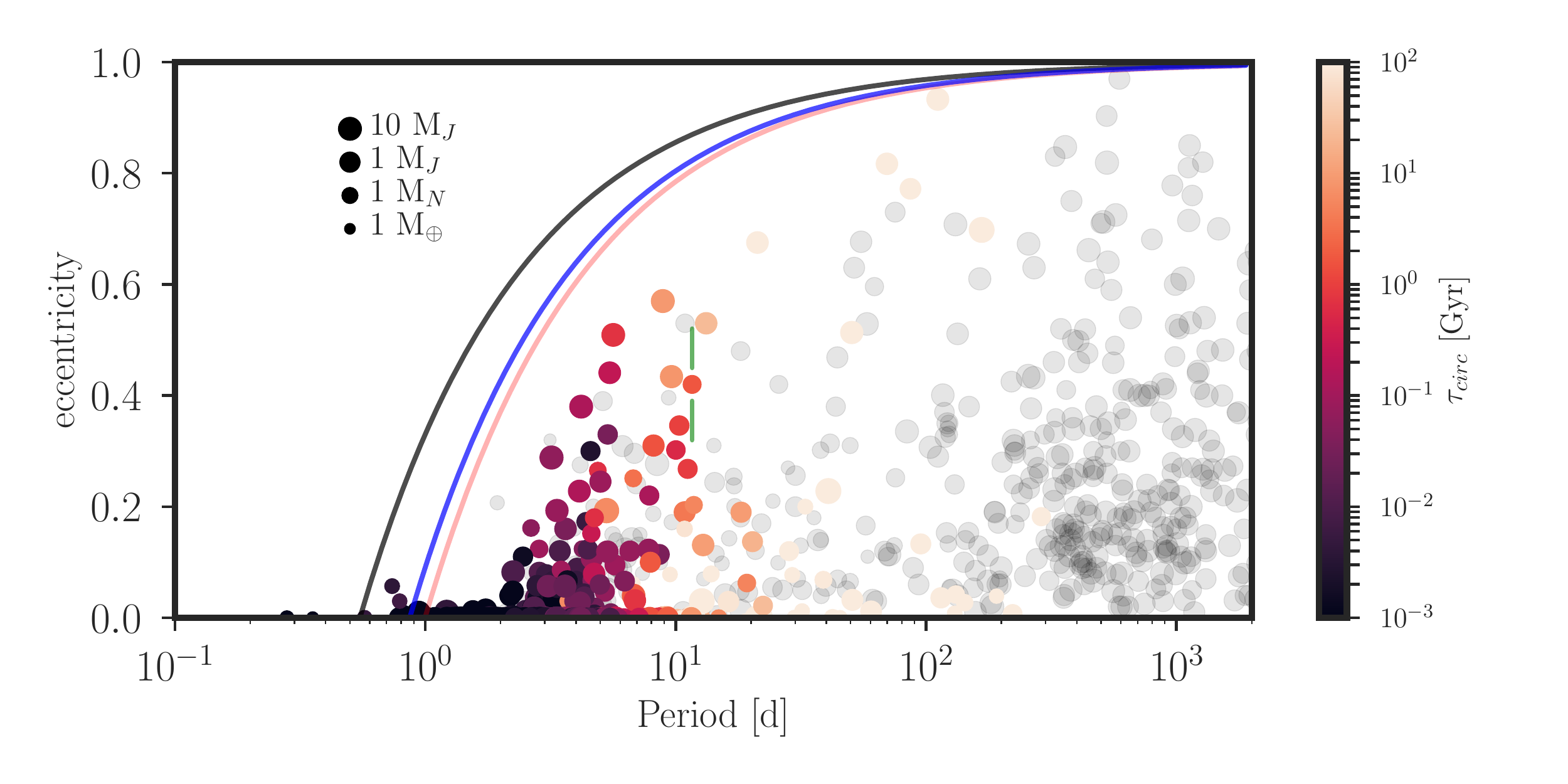}
 \caption{Scatter plot of the eccentricity as a function of the orbital period. Coloured circles correspond to transiting planets where the colour represents the
 tidal circularization timescale. The grey circles correspond to non-transiting
 planets. The size of the circles represents the mass of the object (\mpl$\sin{i}$ in the case of non-transiting planets). \plname can be identified as the only system having green error bars. The three solid lines correspond to the
 maximum eccentricity that an orbit can take such that the periapsis distance lies
outside the Roche limit. The red line corresponds to the limit for the physical parameters of the \stname\ system. The blue (black) line corresponds to the limit in the case of a Neptune-like (Jupiter-like) planet orbiting a sun-like star. \plname\
is among the most eccentric systems having $P<50$\,d. Additionally, \plname\ has one of
the shortest circularization timescales among this high eccentricity systems.}
 \label{ecc}
\end{figure*}

\subsection{Structure}

Figure \ref{m-r} displays the full population of well characterized
transiting planets where the super Neptune region is highlighted in green.
\plname\ joins this sparsely populated group of systems
having masses in the $0.054\,\mjup < \mpl < 0.18\,\mjup$ range.
According to TEPcat \citep{tepcat}, there are another 25 systems in this region
of the parameter space with radii and masses measured at the
25\% level. Among these systems, \plname\ shares similar structural properties to HATS-8b \citep{hats-8b}, WASP-107b \citep{wasp-107b}, and WASP-139b \citep{wasp-139b}. These four systems have particularly low densities.

A clear property of the super Neptune population is that it covers
a wide range of internal structures, as can be identified from the
different values that the radii can take. In the case of low irradiated
planets \citep[$\teq < 1000$\,K,][]{kovacs:2010} the radius is mostly set by the
amount of solid material in the interior. In this context, \plname\ and the
other low density super Neptunes seem to be depleted in solid material when compared to the rest of the population. Specifically, \citet{thorngren:2016} inferred the amount of solid material for some of these planets, finding that the fraction of solids for WASP-139b is consistent with 40\%, while the more compact systems (GJ436b, HAT-P-11b, K2-27b) tend to have values between 70\% and 90\%. \stname\ and the host stars of the other low density Neptunes do not seem particularly metal poor ($0.0<\feh<0.3$) which could be linked to a proto-planetary disc with a low concentration of solids. Therefore, the small gas-to-solids ratio in the envelope of these low density planets could be produced by their formation and accretion history.

Even though super Neptune systems show a wide variety of structures and fraction of metals, they should share a similar formation mechanism, in which the embryo doesn't quite enter into the process of run away accretion of the gaseous envelope because it doesn't reach the pebble isolation mass \citep[e.g.][]{ida}.
Nonetheless, these conditions can be satisfied at different regions of the proto-planetary disc. For example, \citet{bitsch} demonstrated that ice giants can be formed at very large orbital distances (40 AU) but also inside 5 AU, while gas giants form in the region in between. The information about the metal content could be an additional variable that further constrains the location where the planet accretes most of its envelope. In this context, the measurement of the atmospheric metallicities and C/O ratios for these planets is highly valuable, because this value can
be used to constrain the region where the planet accreted most of 
its envelope \citep{mordasini:2016, Espinoza:2017}. The measurement of the atmospheric  C/O ratios is possible through transmission spectroscopy, and has been successfully obtained for the low density super Neptune WASP-107b using HST/WFC3 \citep{kreidberg:2018}, where the data is consistent with
a sub solar C/O ratio. \plname\ is a well suited comparison
candidate to search for water molecules and constrain the C/O
ratio of planets in the ice/gas transition range. Its bright
host star, coupled to its low density, translates to a
transmission spectroscopy metric (TSM) of 170, which according
to \citep{tsm}, puts it in the group of highest priority targets
for atmospheric characterization with JWST.

\subsection{Orbital evolution}
Another particular property of \plname\ besides its low density
is  the significant eccentricity of its orbit. Figure \ref{ecc}
shows that \plname\ is among the most eccentric planets having
orbital periods shorter than 50d.
In the absence of external mechanisms, the evolution of the orbit
of \plname\ should be dominated by tidal interactions produced by
the star onto the planet during periastron passages.
In this way \plname\ should circularize its orbit while migrating
closer to the star \citep{jackson:2008}.
As can be also identified in Figure \ref{ecc}, \plname\ presents
one of the shortest circularization timescales ($\taucirc=1.7$\,Gyr)
of the population transiting systems with $P<50$~d and $e>0.4$, which
is principally caused by the relatively low mass of \plname.
For example, CoRoT-10b \citep{corot-10b} and HD17156 \citep{hd17156}
with a masses of \mpl = 2.76 \mjup\ and \mpl = 3.3\mjup, respectively, present circularization timescales greater than 20 Gyr.
Two eccentric systems of this sample that present short
circularization timescales are HAT-P-2b \citep[\mpl = 8.87\mjup,  $\taucirc=0.7$\,Gyr,][]{hat-p-2b}, and HAT-P-34b \citep[$\mpl = 3.33\,\mjup$,  $\taucirc=0.3$\,Gyr,][]{hat-p-34b}.
Both systems, however, are significantly younger ($\sim 2$\,Gyr) than \stname\ (8.5\,Gyr), which could help in explaining their non-circular orbits. 

\begin{figure}
 \includegraphics[width=\columnwidth]{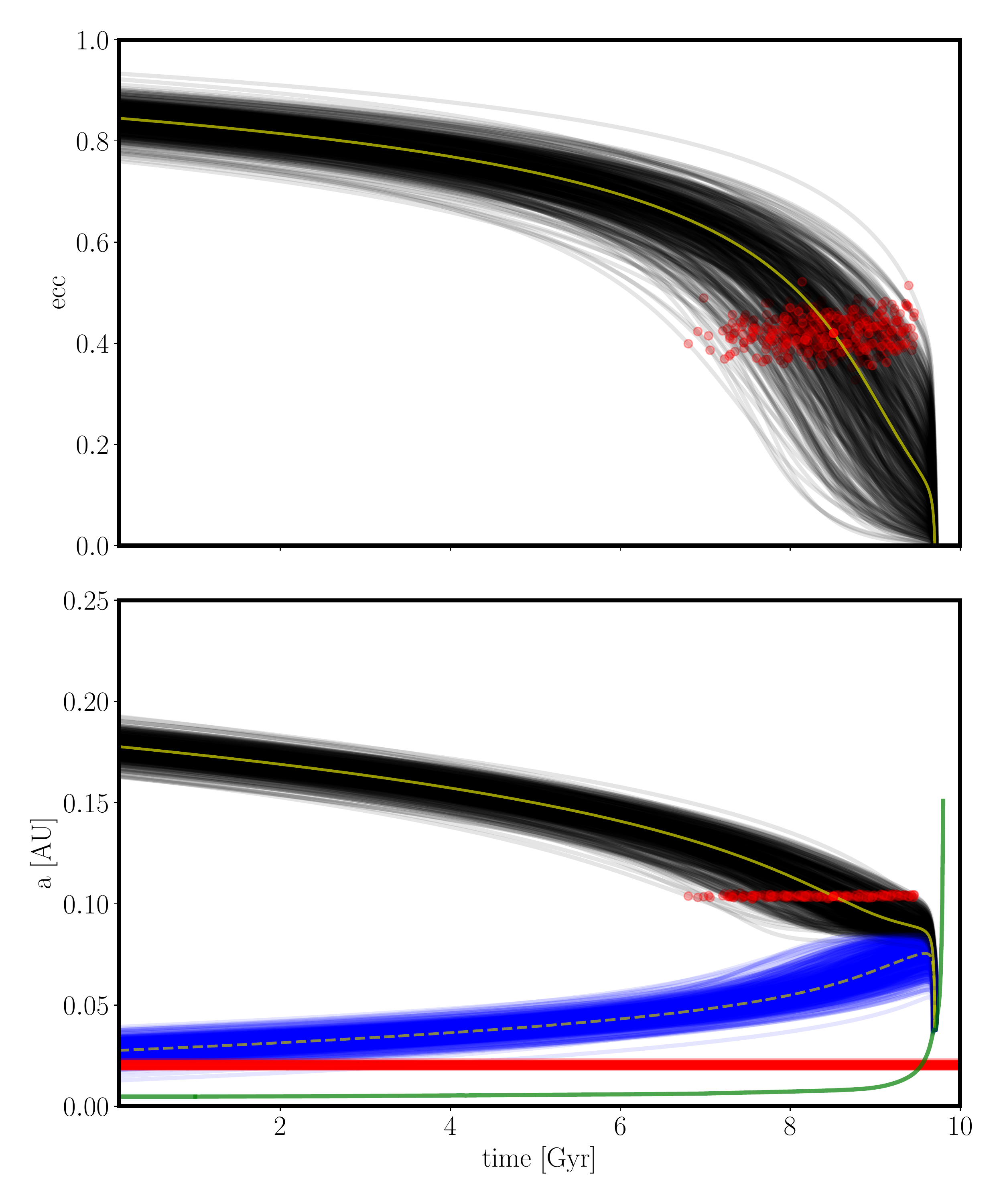}
 \caption{Results obtained for the simulations of the orbital evolution
 of \plname\ under the influence of the tidal interactions with the star.
 The top panel corresponds to the eccentricity evolution, where the black yellow
 line corresponds to the evolution given by the adopted current value of the eccentricity ($e=0.42$). The middle panel corresponds to the evolution of
 the semi-major axis (grey lines), periapsis distance (blue lines),
 Roche limit (red lines), and stellar radius (green line).
 }
 \label{tidal1}
\end{figure}

In order to check if the derived orbital and physical parameters
translate in a feasible scenario for the existence of the \plname\,
we studied the tidal evolution history of the system.
We used the equations presented in \citet{jackson:2009} that govern
the change in $a$ and $e$ as a function of time affected by the tidal interactions produced by the star on the planet and vice-versa.
We also consider the changes in the radius of the host star as it
leaves the main sequence by generating an interpolated evolutionary
track from the Yonsei-Yale isochrones.
Figure \ref{tidal1} shows the evolution of $a$ and $e$ for the
parameters that we obtained for the \stname\ system, where we assume
a tidal quality factor of $Q_{\star}=10^5$ \citep{jackson:2009} for
the star, and a Jupiter-like $Q_\mathrm{P}=3\times10^4$ for the planet \citep{lainey:2009}.

We performed several simulations of the tidal evolution by selecting stellar and planetary values from random distributions using the
outcomes of our joint modelling.
We obtain that \plname\ will be engulfed by its evolving host star before circularizing its orbit in $\approx 1$\,Gyr from now.
More interestingly, we find that if we go back in time, the planet pericentre distance gets very close to the roche limit of the system, but does not cross it. For example, when the system was 1\,Gyr old,
a semi-major axis of $a=0.175$\,AU and an eccentricity of $e=0.85$, translates into a pericentre distance of just $1.4$ Roche radii. In this way, one simple origin for the current parameters of the \stname\ system is that \plname\ could have formed beyond 0.2 AU, and then migrated though the proto-planetary disc to $\approx 0.2$\,AU \citep[e.g.][]{ida2008}. As soon
as the gaseous disc was dispersed ($\sim 100$\,Myr), gravitational interactions with additional objects in the system excited the
eccentricity of \plname\ to $e\approx0.9$ \citep[e.g. ][]{scat} and it has been migrating
through tidal interactions since then.

\section*{Acknowledgments}
R.B.\ acknowledges support from FONDECYT Post-doctoral Fellowship Project No. 3180246.
A.J.\ acknowledges support from FONDECYT project 1171208,
BASAL CATA PFB-06, and project IC120009 ``Millennium Institute
of Astrophysics (MAS)" of the Millennium Science Initiative,
Chilean Ministry of Economy. 
A. Z.\ acknowledges support by CONICYT-PFCHA/Doctorado Nacional-21170536, Chile.
M.R.D.\ acknowledges support by CONICYT-PFCHA/Doctorado Nacional-21140646, Chile.
J.S.J.\ acknowledges support by FONDECYT project 1161218 and partial support by BASAL CATA PFB-06.
This paper includes data collected by the K2 mission. Funding
for the K2 mission is provided by the NASA Science Mission directorate.
This work has made use of data from the European Space Agency (ESA)
mission {\it Gaia} (\url{https://www.cosmos.esa.int/gaia}), processed by
the {\it Gaia} Data Processing and Analysis Consortium (DPAC,
\url{https://www.cosmos.esa.int/web/gaia/dpac/consortium}). Funding
for the DPAC has been provided by national institutions, in particular
the institutions participating in the {\it Gaia} Multilateral Agreement.
Based on observations collected at the European Organisation for Astronomical
Research in the Southern Hemisphere under ESO programmes 094.C-0428(A), 0101.C-0407(A), 0101.C-0497(A).

\bibliographystyle{mnras}
\bibliography{main}

\appendix
\begin{table*}
 \centering
  \caption{Radial velocity and bisector span measurements for \stname.}
  \label{rvsbss}
  \begin{tabular}{@{}lccccr@{}}
  \hline
     BJD              &   RV                  & $\sigma_{RV}$ &   BIS               &  $\sigma_{BIS}$ &  Instrument \\
    (-2400000)     &   [km s$^{-1}$]  & [km s$^{-1}$]   &   [km s$^{-1}$] &   [km s$^{-1}$] &   \\
 \hline

58145.8228641 & 3.3322 & 0.0090 & -0.024 & 0.010 & Coralie \\
58167.7218311 & 3.3350 & 0.0070 & -0.000 & 0.009 & FEROS \\
58198.7443322 & 3.3045 & 0.0090 & -0.017 & 0.010 & Coralie \\	
58171.7144194 & 3.3426 & 0.0020 & -0.002 & 0.002 & HARPS \\
58177.7335895 & 3.3382 & 0.0055 & -0.006 & 0.007 & HARPS \\
58178.7258003 & 3.3352 & 0.0059 &  0.001 & 0.008 & HARPS \\
58199.6738848 & 3.3195 & 0.0090 &  0.003 & 0.010 & Coralie \\
58211.7074500 & 3.3267 & 0.0020 &  0.009 & 0.002 & HARPS \\
58212.6562454 & 3.3359 & 0.0020 &  0.005 & 0.002 & HARPS \\
58213.6244947 & 3.3432 & 0.0020 &  0.006 & 0.002 & HARPS \\
58214.6231712 & 3.3437 & 0.0028 &  0.007 & 0.004 & HARPS \\
58235.5508853 & 3.3301 & 0.0055 & -0.006 & 0.007 & HARPS \\
58236.4894689 & 3.3460 & 0.0022 &  0.008 & 0.003 & HARPS \\
58239.5556333 & 3.3285 & 0.0070 & -0.006 & 0.009 & FEROS \\
58239.5677837 & 3.3214 & 0.0070 & -0.008 & 0.009 & FEROS \\
58239.5796199 & 3.3283 & 0.0070 & -0.019 & 0.009 & FEROS \\
58241.7013063 & 3.3205 & 0.0070 &  0.009 & 0.009 & FEROS \\
58241.6923621 & 3.3269 & 0.0070 &  0.001 & 0.009 & FEROS \\
58242.6012465 & 3.3304 & 0.0070 & -0.012 & 0.009 & FEROS \\
58242.6088207 & 3.3407 & 0.0070 & -0.003 & 0.009 & FEROS \\
58243.6728970 & 3.3124 & 0.0070 & -0.004 & 0.009 & FEROS \\
58243.5960625 & 3.3170 & 0.0070 &  0.001 & 0.010 & FEROS \\
58244.5790597 & 3.3036 & 0.0070 & -0.009 & 0.009 & FEROS \\
58244.6109274 & 3.3106 & 0.0070 & -0.003 & 0.010 & FEROS \\
58249.6027837 & 3.3225 & 0.0070 & -0.009 & 0.010 & FEROS \\
58249.4650452 & 3.3289 & 0.0074 & -0.025 & 0.011 & FEROS \\
58250.5395590 & 3.3259 & 0.0070 & -0.005 & 0.009 & FEROS \\
58251.5932739 & 3.3345 & 0.0070 & -0.006 & 0.009 & FEROS \\
58261.5307685 & 3.3176 & 0.0070 & -0.015 & 0.009 & FEROS \\
58261.5458842 & 3.3173 & 0.0070 &  0.000 & 0.009 & FEROS \\
58261.5383175 & 3.3179 & 0.0070 & -0.020 & 0.009 & FEROS \\
58262.4795489 & 3.3458 & 0.0045 & -0.002 & 0.006 & HARPS \\
58262.5348530 & 3.3153 & 0.0070 & -0.010 & 0.009 & FEROS \\
58262.5424179 & 3.3121 & 0.0070 & -0.018 & 0.009 & FEROS \\
58262.5499684 & 3.3134 & 0.0070 & -0.004 & 0.009 & FEROS \\
58263.5071991 & 3.3407 & 0.0070 & -0.017 & 0.009 & FEROS \\
58263.5147548 & 3.3325 & 0.0070 & -0.010 & 0.009 & FEROS \\
58263.5222990 & 3.3409 & 0.0070 & -0.010 & 0.009 & FEROS \\
58263.5298571 & 3.3287 & 0.0070 & -0.012 & 0.009 & FEROS \\
58263.5435131 & 3.3528 & 0.0020 &  0.001 & 0.002 & HARPS \\
58264.5319999 & 3.3514 & 0.0052 &  0.006 & 0.007 & HARPS \\
58264.5525416 & 3.3394 & 0.0081 &  0.006 & 0.012 & FEROS \\
58264.5601077 & 3.3248 & 0.0070 & -0.014 & 0.010 & FEROS \\
58264.5676783 & 3.3234 & 0.0070 & -0.019 & 0.010 & FEROS \\
58264.5752260 & 3.3214 & 0.0070 & -0.005 & 0.011 & FEROS \\
58264.5827913 & 3.3387 & 0.0072 &  0.016 & 0.011 & FEROS \\
58265.5360829 & 3.3327 & 0.0070 & -0.020 & 0.010 & FEROS \\
58265.5436332 & 3.3301 & 0.0070 & -0.011 & 0.009 & FEROS \\
58265.5511918 & 3.3265 & 0.0070 & -0.007 & 0.009 & FEROS \\
58265.5587443 & 3.3377 & 0.0070 &  0.003 & 0.009 & FEROS \\
58265.5663100 & 3.3299 & 0.0070 & -0.005 & 0.009 & FEROS \\
58266.4923891 & 3.3277 & 0.0077 &  0.015 & 0.012 & FEROS \\
58266.4999270 & 3.3361 & 0.0070 & -0.039 & 0.011 & FEROS \\
58266.5059263 & 3.3531 & 0.0048 &  0.007 & 0.006 & HARPS \\
58266.5074907 & 3.3264 & 0.0070 & -0.011 & 0.010 & FEROS \\
58266.5150622 & 3.3402 & 0.0070 & -0.002 & 0.010 & FEROS \\
58266.5226106 & 3.3360 & 0.0070 & -0.012 & 0.010 & FEROS \\
\hline

\end{tabular}
\end{table*}

\label{lastpage}

\end{document}